\documentclass[twocolumn,a4paper]{aa}
\usepackage{graphicx}
\usepackage{aalongtable}
\input{3157null.defs}
\usepackage[sort&compress]{natbib}
\bibpunct{(}{)}{:}{a}{}{,} 
\usepackage{txfonts}
\newcommand{\be}{\begin{eqnarray}}
\newcommand{\ee}{\end{eqnarray}}

\begin{document}

\title{Simultaneous Single-Pulse Observations of Radio Pulsars:
V. On the Broadband Nature of The Pulse Nulling Phenomenon in 
PSR B1133+16} 

\author{N. D. R. Bhat \inst{1} \and Y. Gupta \inst{2} \and 
M. Kramer \inst{3} \and A. Karastergiou \inst{4} \and 
A. G. Lyne \inst{3} \and S. Johnston \inst{5} }

\institute{Centre for Astrophysics and Supercomputing, Swinburne University of Technology, 
Hawthorn, Vic 3122, Australia \and 
National Centre for Radio Astrophysics, Tata Institute of Fundamental Research, Ganeshkhind, 
Pune 411007, India \and
Jodrell Bank Observatory, University of Manchester, Macclesfield, Cheshire, SK11 9DL, UK \and
IRAM, 300 rue de la Piscine, Domaine Universitaire 38406 Saint Martin d'Hères, France \and 
Australia Telescope National Facility, CSIRO, PO Box 76, Epping, NSW 1710, Australia }


\bigskip

\date{Accepted 19 October 2006}

\abstract 
{} 
{In this paper we revisit the well-known phenomenon of pulse nulling using 
high-quality single-pulse data of PSR B1133+16 from simultaneous multifrequency 
observations.}  
{Observations were made at 325, 610, 1400 and 4850 MHz as part of a joint 
program between the European Pulsar Network (EPN) and the Giant Metrewave 
Radio Telescope (GMRT). The pulse energy time series are analysed to derive 
improved statistics of nulling pulses as well as to investigate the frequency 
dependence of the phenomenon.} 
{The pulsar is observed to be in null state for approximately 15\% of the time; 
however, we find that nulling does not always occur simultaneously at all four
frequencies of observation. We characterise the statistics of such ``selective 
nulling'' as a function of frequency, separation in frequency, and combination 
of frequencies. The most remarkable case of such selective nulling seen in our 
data is a significantly large number of nulls ($\approx$6\%) at lower frequencies, 
that are marked by the presence of a fairly narrow emission feature at the highest 
frequency of 4850 MHz.  We refer to these as ``low frequency (LF) nulls.'' We 
characterise the properties of high frequency (HF) emission at the occurrence 
of LF nulls, and compare and contrast them with that of ``normal emission'' at 
4850 MHz. Our analysis shows that this high frequency emission tends to occur 
preferentially over a narrow range in longitude and with pulse widths typically 
of the order of a few milliseconds. 
We discuss the implications of our results for the pulsar emission mechanism 
in general and for the broadbandness of nulling phenomenon in particular. 
Our results signify the presence of an additional process of emission which does 
not turn off when the pulsar nulls at low frequencies, and becomes more prominent 
at higher frequencies.
Our analysis also hints at a possible outer gap origin for this new population of pulses, 
and thus a likely connection to some high-energy emission processes that occur in the 
outer parts of the pulsar magnetosphere.}  
{}

\keywords{pulsars: general -- pulsars: individual, PSR B1133+16 -- ISM: general -- radiation
mechanism: non-thermal}

\titlerunning{Simultaneous Single-Pulse Observations}

\authorrunning{Bhat et al.}

\maketitle

\section{Introduction}\label{s:intro}

\bigskip

Radio emission from pulsars is known to vary on a wide range of
time scales, from as short as one pulse period to many hours or
days. Many pulsars are known to exhibit the phenomenon of ``nulling,"
where the emission appears to cease, or is greatly diminished, for a
certain number of pulse periods. Typical time scales of nulling are of
the order of a few pulse periods, however it may last for up to many 
hours in certain pulsars; for example, PSR B0826--34 which is active 
for only $\sim$20\% of the time \citep{durdin1979}.

\medskip

Ever since its discovery \citep{backer1970} and early investigations
\citep{ritchings1976}, the phenomenon of nulling has remained as a
vital clue for understanding the elusive pulsar emission mechanism. 
Several authors have investigated this phenomenon in detail (e.g., 
\citet{rankin1986,biggs1992}), and it is fairly well established that
(i) the phenomenon is intrinsic to the pulsar, (ii) it is possibly 
broadband, and (iii) the null fraction (NF) is strongly correlated 
to the pulse period \citep{ritchings1976,biggs1992}. It is also known 
that the null fraction (NF) depends on the pulsar class, and that 
single-core pulsars have rather low values of NF \citep{rankin1986}. 
However, the suggested correlation between the null fraction and age 
is not strongly supported within a given pulsar class. As is the case 
with several other phenomena of pulsar signals, a satisfactory explanation 
for nulling, particularly the physics that governs it, remains largely 
elusive.

\medskip

Studies of individual pulses from simultaneous multifrequency
observations provide valuable insights into the pulsar emission
mechanism. In particular, such studies help address the broadband
nature and frequency dependence of the intrinsic phenomena such as
drifting and nulling. Such studies however require coordinated
multi-station observations which are hard to realise in
practice. Consequently, few such studies have been reported in the
past, and most of these were made in the early days of pulsar research
\citep{robinson1968,backer1974,davies1984}. Moreover, most of these
experiments involved observations at only 2 to 3 frequencies, and
focussed primarily on lower frequencies ($\la$ 2.6 GHz).  These
observations led to a general understanding that phenomena such as
drifting, nulling and moding are generally of broadband nature at low
observing frequencies.

\medskip

This work is fifth in a series of papers describing simultaneous
multifrequency observations of radio pulsars. Observations were made
in January 2000 as part of a joint collaborative effort between the
European Pulsar Network (EPN; \citet{lorimer1998}) and the Giant
Metrewave Radio Telescope (GMRT) in India. These observations 
led to long stretches of
high quality data for five pulsars over a wide frequency range, 
from 0.24 to 4.85 GHz \citep{bhat2001}. Data at 1.4 and 4.85 GHz
(from the Lovell and Effelsberg telescopes respectively) were recorded
in full polarisation, while data at lower frequencies (from the
GMRT) were recorded in total power. The science goals included
investigation of a variety of phenomena related to pulsar emission.
The earlier papers in this series focussed on polarisation and spectra
of individual pulses (\cite{karas2001,karas2002,karas2003,kramer2003};
hereafter Paper IV). 
In this paper we
present an in-depth analysis of the pulse nulling phenomenon using data at
4 frequencies for PSR B1133+16, with particular emphasis on
investigating the broadband nature of the phenomenon.

The remainder of the paper is organised as follows. In \S 2 we describe the
details of observations and data reduction. Pulse energy time series are 
presented in \S 3, and we discuss the statistics of nulling in \S 4. In later
sections we focus our attention on some key results from our analysis,
most notable of which is evidence for a ``selective nulling" phenomenon
that is characterised by mostly single-peaked, narrow emission features 
at the highest frequency, 4850 MHz, and nulls at lower frequencies.
Our conclusions are presented in \S~\ref{s:conc}.

\section{Simultaneous Multifrequency Observations}\label{s:simobs}

The data presented in this paper was obtained from simultaneous
observations at four different frequencies, viz. 325, 610, 1400 and 
4850 MHz. Data at the higher two frequencies were recorded over 
bandwidths of 500 MHz (Effelsberg) and 32 MHz (Lovell), while the 
data-taking system at the GMRT employed a bandwidth of 16 MHz. 
The observations using Effelsberg and Lovell have been described 
by \citet{karas2002}. Observations at the GMRT capitalised on the 
instrument's unique capability whereby observations at more than 
one frequency can be done simultaneously, by configuring the 
telescope in a multiple sub-array mode \citep{gupta2002}. In this
mode, data at the two frequencies were recorded by the same data 
logging system, where the signals from both the frequencies were 
added together after square-law detection, and the pulsar dispersion 
delay was used to separate the two data streams in the off-line 
analysis. Data from all telescopes were stored for off-line processing.

At Effelsberg and Lovell, data were recorded as a single,
uninterrupted session, whereas data-taking at the GMRT was split 
into two sessions (of durations 1932 and 2441 pulse periods, i.e., 
in total 4373 pulse periods) due to maximum file size limitations 
for a single scan.  For the convenience of analysis, 
as well as to allow useful cross-checks of any statistical analyses 
that we perform, we treat data from the two sessions as two separate 
data sets, and hereafter refer to them as \dataa and \datab, respectively. 
Data from
all three telescopes were converted to a common EPN format and
time-aligned following the procedures detailed in \citet{karas2001}.
Further, data for all four frequencies were smoothed to an effective
time resolution of 1159.8 $\mu$s (off-line dispersion was performed on
data from the GMRT, while data at Lovell were recorded after
correcting for dispersion on-line)\footnote{The dispersion effect is
negligible at the highest frequency 4850 MHz, and hence no correction
was applied for data from Effelsberg.}. A small fraction of the data
(roughly 9\%), particularly at the GMRT frequencies, was affected by
radio frequency interference (RFI), and is hence excluded from subsequent 
analyses.

As described in Paper IV, flux density calibration was 
performed at all four frequencies. At Effelsberg and Lovell, the pulsar data 
was compared to that of a pre-calibrated noise diode, which itself
was compared to the strength of known flux calibrators observed 
during pointing observations before, after and during the observations.
The estimated uncertainty resulting from this calibration
scheme is about 10\%. At the GMRT, known flux calibrators such as 3C147,
3C295 and 3C286 were observed before and after every pulsar observation.
Unlike the pulsar data, the calibration data were gathered separately 
at two frequencies, i.e., one frequency at a time, using the summed 
signal from only those antennas that were selected for the frequency.
From these data, an effective calibration scale was computed for each
of the two frequencies. Using the nearest available calibration source, 
this scale was then applied to the ``on--off'' deflection of the pulsar, 
separately for each frequency of observation, in order to flux 
calibrate the pulsar signal at both frequencies. As demonstrated in
Paper IV, the average flux densities estimated from our calibration
scheme are consistent with what is known from literature at the
frequencies of observation.

\section{Pulse Energy Time Series}\label{s:epul}

\subsection{Detection of null and on states}

Conventionally, a null state signifies our inability to detect pulsar emission 
over one or more pulse periods. This {\it inability} depends on the sensitivity 
of the observing system, and consequently, sensitive single-pulse observations 
are ideal for accurate characterisation of pulsar nulling properties.  
For PSR B1133+16, typical flux densities at our observing frequencies are 
$\sim$20 to $\sim$100 times larger than the minimum detectable flux densities 
obtained from our observing parameters. Such high quality data allow an 
in-depth study of the nulling behaviour of this pulsar.

\begin{figure}
\resizebox{\hsize}{!}{\includegraphics[angle=0]{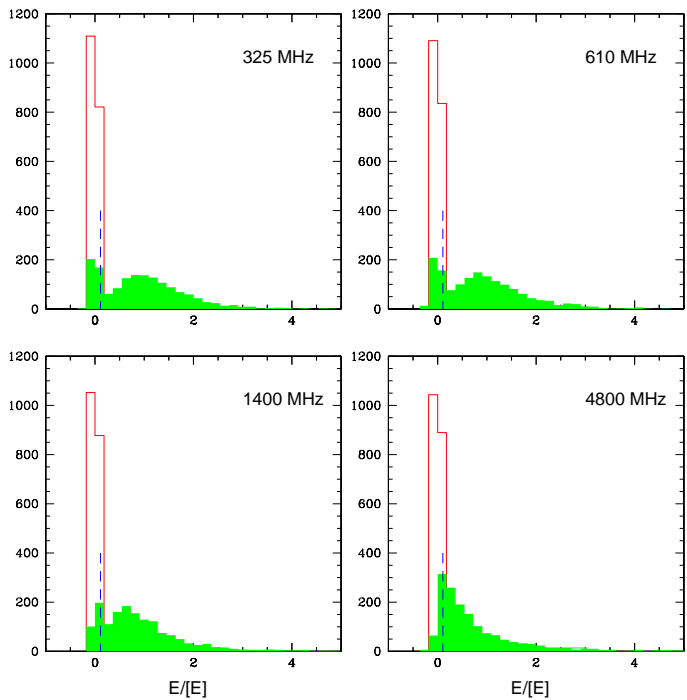}}
\caption{ON and OFF pulse energy histograms (shown as shaded and unshaded regions 
respectively) for the \dataa at four frequencies of observation, 325, 610, 1400 
and 4850 MHz. The pulse energy E is normalised with its running mean over 200 pulse 
periods, [E]. The vertical dashed line is the mean of 3-$\sigma$ pulse energy 
thresholds that are used to define the state of the pulse.}
\label{fig:nullhist}
\end{figure}

Traditionally, the occurrence of nulling in a pulsar is characterised by the pulse
nulling statistics that are derived from pulse energy distributions (e.g. Biggs 1992; 
Ritchings 1976).  ON and OFF pulse energy histograms for our four frequency 
observations are shown in Fig.~\ref{fig:nullhist}.  At lower frequencies, 
we can see a classical separation of the ON pulse histograms into two components, 
with a zero energy excess that signifies the presence of nulling.  At the higher two
frequencies, the distributions tend to merge with each other, as the signal-to-noise
ratio (S/N) for the pulses is, on the whole, poorer.  

Using these distributions and a method similar to that described by Ritchings (1976),
we have estimated the null fraction (NF) values for the two data sets 
(Table~\ref{tab:alpha}), at the two lower frequencies. 
The method does not yield meaningful estimates at the higher frequencies 1400 and 4850 MHz, 
where the ON and OFF distributions tend to merge. The lower and upper limits obtained in 
this manner are denoted as \alphalow and \alphahigh respectively. Values for the NF at 
a given frequency are quite consistent between the two data sets.  Furthermore, they are 
consistent with values reported (at 408 MHz) in the literature for this pulsar (Biggs 1992; 
Ritchings 1976).

Whereas pulse energy distributions provide a good tool for estimating statistical 
values related to the nulling phenomenon, we need a different approach to identify
individual null pulses at each frequency of observation.  We do this by comparing the
ON pulse energy estimate with a threshold based on the system noise level.
The uncertainty in the pulse energy estimate \sigep
is given by $\sqrt {\non} \sigoff$, where \non is the number of ON pulse bins
and \sigoff is the rms of the OFF pulse region.  Using this as a threshold, 
we define pulses with ON pulse energy smaller than $ 3 \times \sigep $ as null pulses.

\begin{figure*}
\centering
{\includegraphics[width=10cm, angle=270]{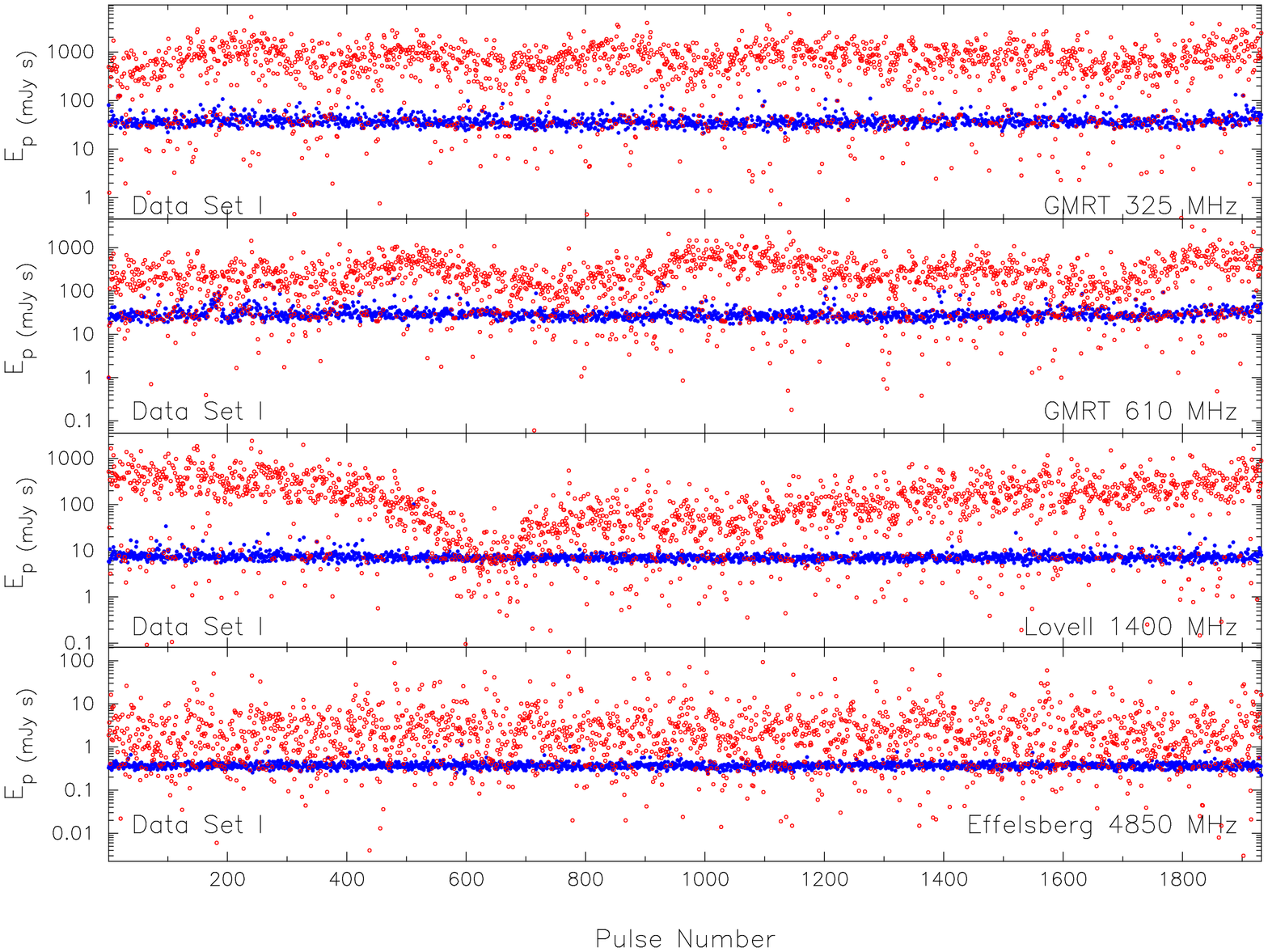}}
{\includegraphics[width=10cm, angle=270]{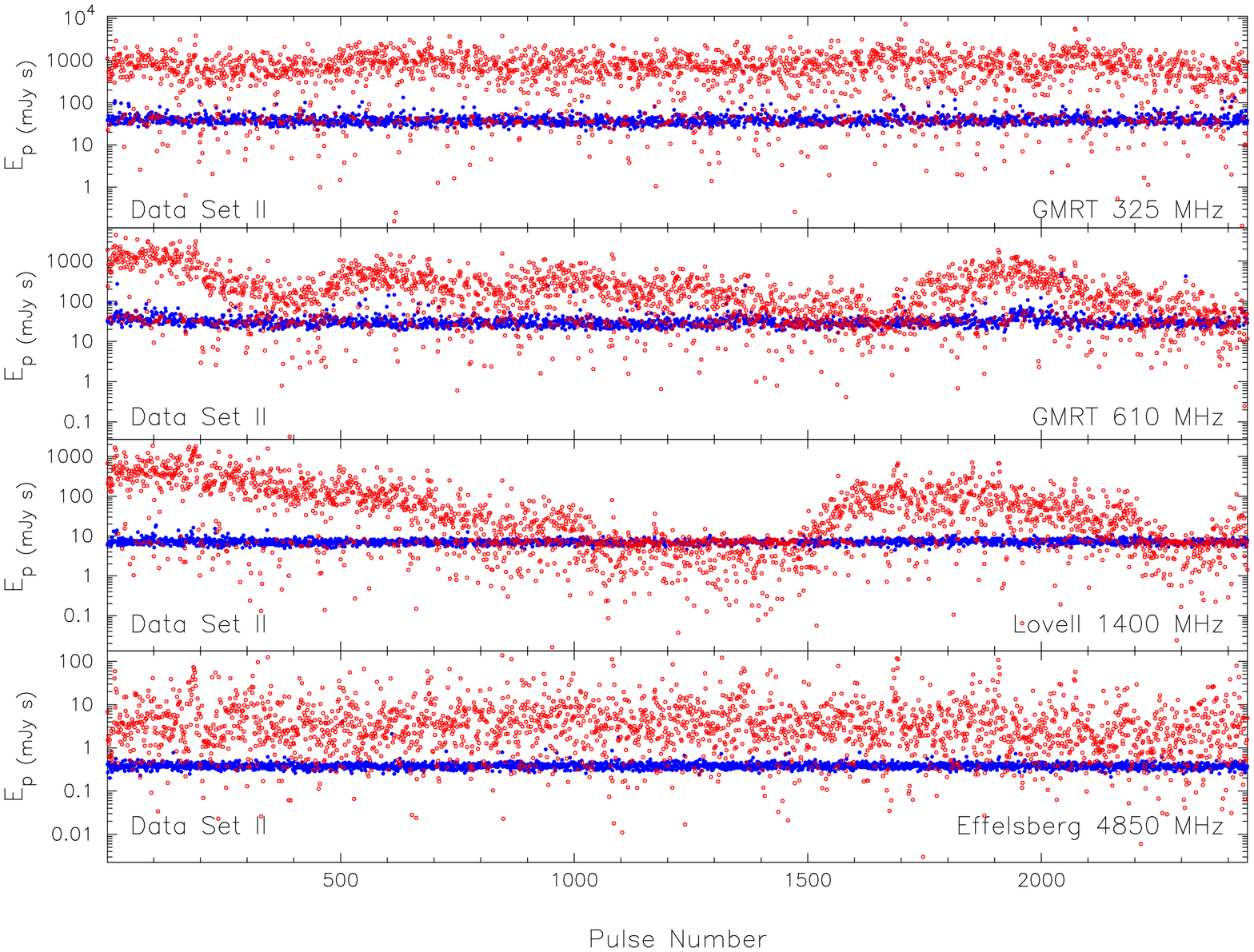}}
\caption{Pulse energy time series of PSR B1133+16 at 325, 610, 1400 and 4850 MHz
(top to bottom in each panel, with {\it top} panel for the \dataa and {\it bottom} 
panel for the \datab).
The blue band of filled circles represents the thresholds below which a pulse is 
considered to be in null state. For 325, 610 and 1400 MHz, energy estimates that 
are below zero are given a fixed value of 1 in the plots. The energy estimates 
are plotted on logarithmic scales for easy identification of null pulses.}
\label{fig:epul}
\end{figure*}

Results from this classification are illustrated in Fig.~\ref{fig:epul} which shows
the time series of pulse energies (on a logarithmic scale for clarity). 
The blue band of filled circles denotes the 3-$\sigma$ threshold.  As can be seen, the
single pulse energies vary by large amounts.  Some of the slower time scale 
variations are likely due to interstellar scintillations (ISS), and this effect 
is discussed in detail in Appendix I.  At each frequency, a significant number of 
individual null pulses can easily be detected.  


\begin{table}
\caption{Estimates of the null fraction at 325 and 610 MHz$^{\mathrm{\dagger}}$} 
\label{tab:alpha} 
\begin{tabular}{cccccc}
\hline\hline
{ } & \multicolumn{2} {c} { Data set I } && \multicolumn{2} {c} { Data set II } \\
\cline {2-3} \cline {5-6} 
Frequency (MHz) & \alphalow & \alphahigh && \alphalow & \alphahigh \\
\hline
325 & 15\%   & 22.5\% &&	15.5\% & 24.5\% \\
610 & 12.5\% & 23\%   &&	14.5\% & 24\%  \\ 
\hline\hline
\end{tabular}                                                        
\begin{list}{}{}
\item[$^{\mathrm{\dagger}}$]Not obtainable at 1400 and 4850 MHz (see text for justification).
\item[] Note: The null fraction at 408 MHz $\sim$ 15$\pm$3\% \citep{ritchings1976}.
\end{list}
\end{table}                                                        

\begin{table*}
\caption{Summary of null counts and exclusive nulls for various frequency combinations} 
\label{tab:nf} 
\begin{tabular}{lccccc}
\hline\hline
{} &
\multicolumn{2} {c} {Data Set I (1932 pulses)$^{\mathrm{a}}$} & {} &
\multicolumn{2} {c} {Data Set II (2441 pulses)$^{\mathrm{a}}$} \\
\cline{2-3} \cline{5-6} \\
{Null Class} & 
{Null Count} &
{Exclusive Nulls} &
{} & {Null Count} &
{Exclusive Nulls} \\
{(1)} & {(2)} & {(3)} & 
{} & {(4)} & {(5)} \\
\hline
4-frequency nulls:&	  &         &&           &          \\
325+610+1400+4850&145 (8.3)&	    &&147 (6.6)  &          \\ 
              &           &         &&           &          \\
3-frequency nulls:&	  &	    &&           &          \\ 
325+610+1400  & 207 (11.8)&62 (3.5) &&315 (14.2) &168 (7.6) \\
325+610+4850  &	155 ( 8.8)&10 (0.6) &&152 ( 6.8) &5   (0.2) \\
325+1400+4850 &	152 ( 8.7)&7  (0.4) &&162 ( 7.3) &15  (0.7) \\
610+1400+4850 &	162 ( 9.2)&17 (1.0) &&164 ( 7.4) &17  (0.8) \\
              &           &         &&           &          \\
2-frequency nulls:&	  &	    &&           &          \\ 
325+610	      &	264 (15.1)&47 (2.7) &&354 (15.9) &34  (1.5) \\
325+1400      &	222 (12.7)&8  (0.5) &&348 (15.7) &20  (0.9) \\
325+4850      &	162 ( 9.2)&0  (0.0) &&168 ( 7.6) &1   (0.1) \\
610+1400      &	233 (13.3)&9  (0.5) &&433 (19.5) &101 (4.5) \\
610+4850      &	189 (10.8)&17 (1.0) &&177 ( 8.0) &8   (0.4) \\
1400+4850     &	177 (10.1)&8  (0.5) &&190 ( 8.6) &11  (0.5) \\
              &           &         &&           &          \\
1-frequency nulls:&	  &	    &&           &          \\ 
325           & 296 (16.9)&17 (1.0) &&413 (18.6) &23  (1.0) \\
610           & 334 (19.1)&27 (1.5) &&590 (26.6)$^{\mathrm{b}}$ &110 (4.9) \\
1400          & 290 (16.5)&34 (1.9) &&848 (38.2)$^{\mathrm{b}}$ &369 (16.6)\\
4850          & 279 (15.9)&75 (4.3) &&222 (10.0) &18  (0.8) \\
\hline\hline
\end{tabular}                                                        
\begin{list}{}{}
\item[] Note: The numbers in parentheses are the percentage null counts, and can be considered as rough estimates of the null fraction.
\item[$^{\mathrm{a}}$] The number of bad pulses (excluded from the analysis): 180 and 220 respectively for the \dataa and the \datab (see text).  
\item[$^{\mathrm{b}}$] Estimation of the null count is probably biased by the interstellar scintillation (ISS) effects (see Appendix A for further details).  
\end{list}
\end{table*}                                                        


\subsection{Binary Time Series of Pulse States}\label{s:bin}

\begin{figure*}
\centering
{\includegraphics[width=9cm,angle=270]{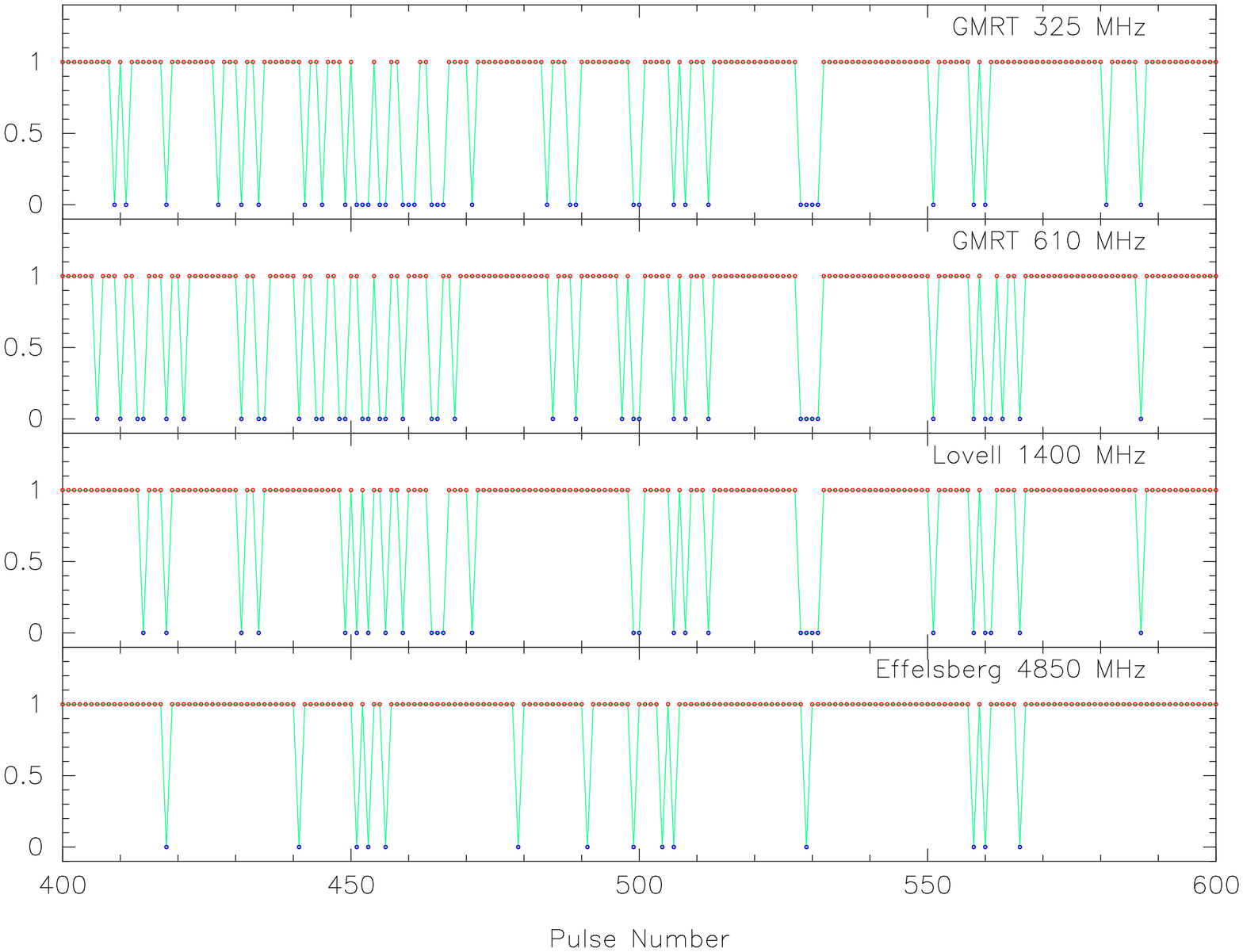}}
\caption{Binary time series of pulse states at four observing frequencies 
325, 610, 1400 and 4850 MHz (top to bottom) for a subset of pulse sequence 
from the \datab. 
The time series is generated for a null threshold of 3\sigep, where \sigep 
is the uncertainty in the pulse energy estimate.}
\label{fig:bin}
\end{figure*}

From the above classification scheme for null pulses, we derive a ``binary time series" 
of pulse states where the ``on" and ``off" (null) states of pulsar emission are assigned 
values of ``1" and ``0", respectively.  An example of such a time series of pulse states 
is shown in Fig.~\ref{fig:bin} where, for clarity, the length of the time series is 
restricted to $\approx$200 pulse periods. This is enough to illustrate the general
trends, and can be summarised as follows:
\begin{enumerate}
\item
To first order, we find that the nulling phenomenon is broadband -- in that most pulses
are in null state simultaneously at all four frequencies -- over a wide range of 
0.3 to 4.9 GHz, and spanning a frequency ratio of $\sim$15, i.e., 4 octaves. 
\item
However, there is a significant number of pulses where nulling does {\it not always} 
occur simultaneously at all frequencies. 
\item
Furthermore, there appears to be a tendency for what we term as ``selective nulling,'' 
wherein the null state is restricted to two or three frequencies of observation (more 
often the lower frequencies), and sometimes just to a single frequency.
\end{enumerate}

To study these aspects more quantitatively, we compute different statistics from
the four frequency binary time series of pulse energies, and investigate their
dependencies on frequency and frequency separation.

\section{Statistics of Null Pulses}\label{s:stat}

The simplest useful statistical quantity is the null count, which is simply the total
number of pulses that are in a null state in the binary time series.  This null count, 
when normalised by the total number of pulses in the data set, can be thought of as 
an estimate of the null fraction.  The usefulness of our data set is that such null 
counts can be estimated for each frequency, as well as for different combinations
of frequencies.  For the latter, only pulses that null simultaneously for the 
desired combination of frequencies are included in the null count. 

Table~\ref{tab:nf} shows a summary of these results, where numbers are tabulated 
separately for data sets I and II.  Column (1) is a listing of the specific class 
(i.e., frequency or the combination of frequencies) for which the null count is 
computed.  Columns (2) and (4) list the number of null pulses (and their percentages)
for a given class. 

\begin{figure}
\centering
{\includegraphics[width=6cm]{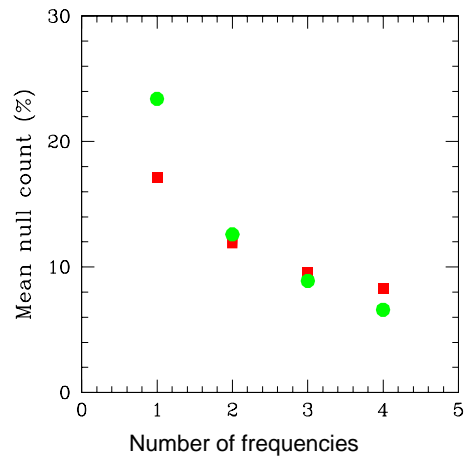}}
\caption{Mean null counts (in percentage) are plotted against the number of observing 
frequencies. This quantity is essentially the mean of all null counts in a given type of null class (see column 1 of Table~\ref{tab:nf}). The red (squares) and green (circles) 
symbols correspond to the data sets I and II respectively. The single-frequency 
point for the \datab is biased upward due to an apparent excess in null counts at 610 
and 1400 MHz for this data set. The number of simultaneous nulls tends to decrease with 
the frequency coverage of the data.} 
\label{fig:nullmean}
\end{figure}

The first thing to note about the results in this table is that our single frequency 
null counts are fairly consistent with our estimates of null fraction in Table~\ref{tab:alpha}, 
with the exception of the 610 MHz estimate for \datab which is biased by the effect of ISS 
(see Appendix A).
Furthermore, the null counts generally appear to reduce systematically as one goes from
the single frequency counts to those for two or more frequency combinations, suggesting 
that the nulling phenomenon is somewhat frequency dependent.  
Thus, for example, the single frequency data show a null fraction of typically $\sim$16\%, 
while the estimate is approximately half this value for the combination of all observing 
frequencies. Furthermore, a scrutiny of the numbers in columns (2) and (4) also reveals 
a tendency for the null fraction to decrease with the frequency coverage of the data. This 
is better illustrated in Fig.~\ref{fig:nullmean} which shows a plot of ``mean null counts''
against the number of frequencies in the null class type under consideration (column 1 
of Table~\ref{tab:nf}). 
Hence, our observations strongly suggest that the nulling phenomenon is {\it not always} 
broadband.  The most significant exceptions to the above trends involve data from 610 and
1400 MHz observations, especially for data set II.  We believe that the large fluctuations
and deep fading of pulse intensities at these frequencies, which lead to increased null 
counts, is due to the phenomenon of ISS.  This is addressed in detail in Appendix A.

\subsection{Nulls and Bursts: Evidence for a Frequency Dependence?}

We digress briefly to address the question of whether the null and burst durations 
have any frequency dependence, as this is something that has never been reported 
before in the literature. Fig.~\ref{fig:bin} hints at such a possibility, with a 
tendency for shorter null durations at higher frequencies.

\begin{figure}
\resizebox{\hsize}{!}{\includegraphics{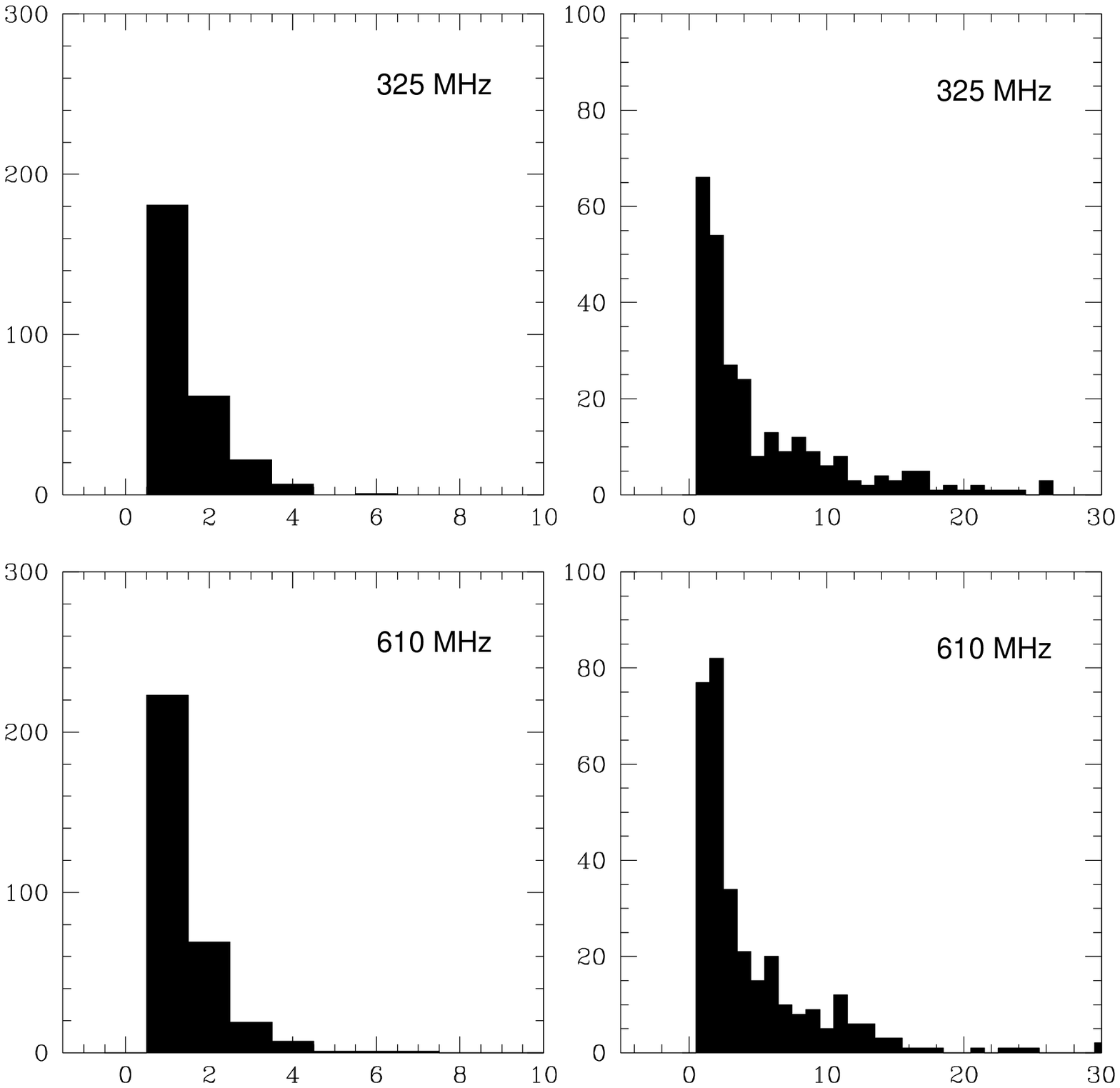}}
\resizebox{\hsize}{!}{\includegraphics{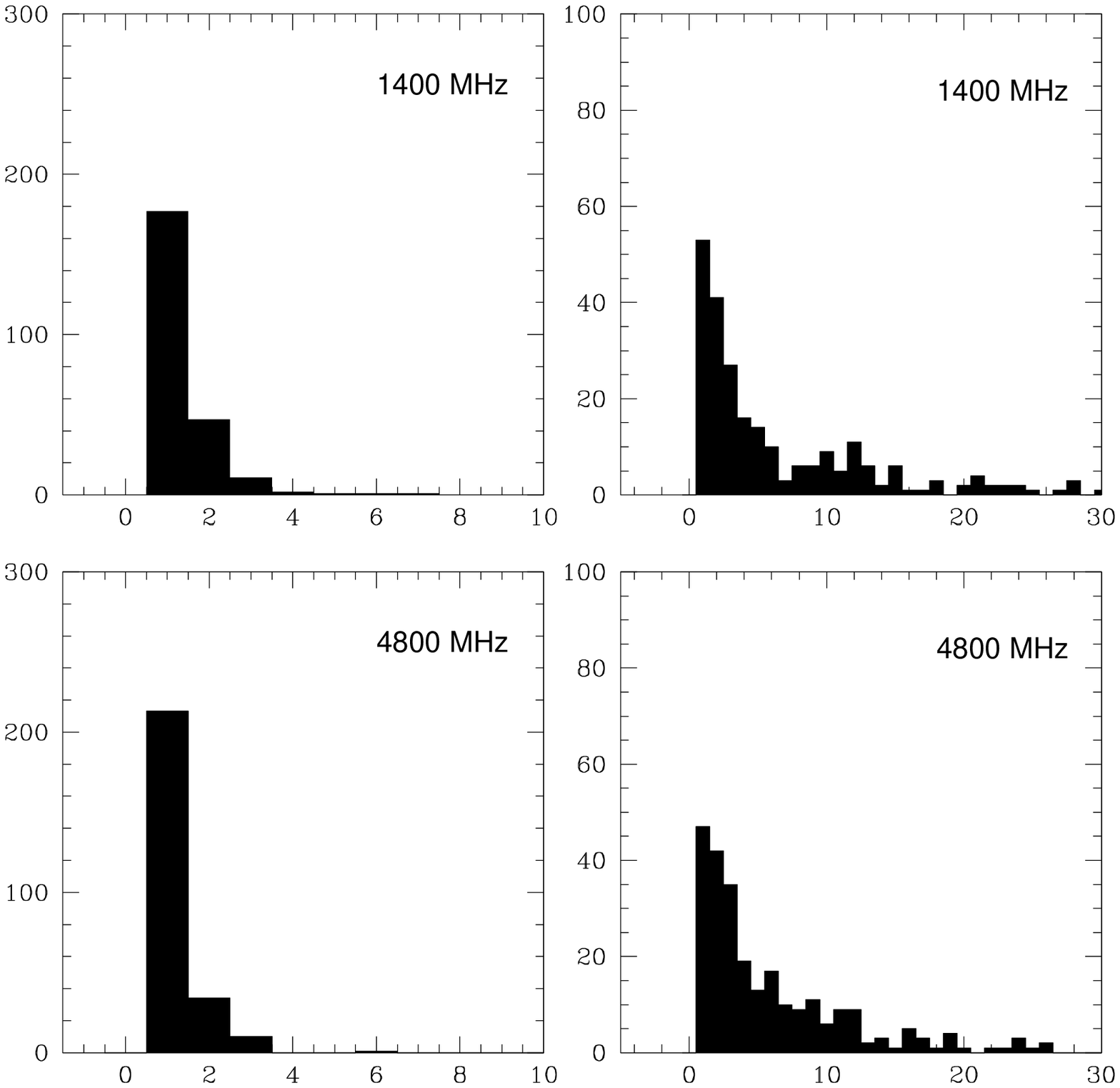}}
\caption{Histograms of null and burst lengths (left and right panels respectively) 
constructed from the \dataa at the four observing frequencies, 325, 610, 1400 and 
4850 MHz. Similar results are also seen for the \datab.}
\label{fig:nullburst}
\end{figure}

\begin{figure}[b]
\centering
{\includegraphics[width=6cm]{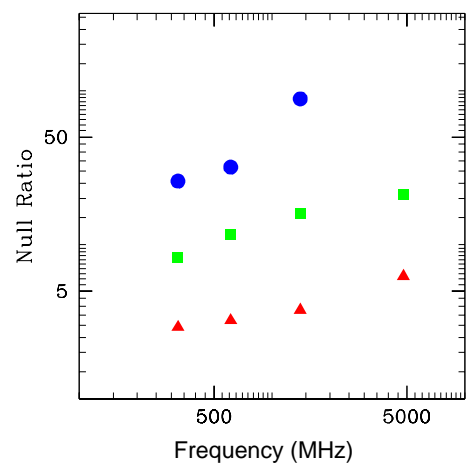}}
\caption{Plot of ``null ratios" as a function of observing frequency. The symbols triangles, 
squares and circles denote the ratios of the number of single-period nulls to the number of 
two-period, three-period and four-period nulls, respectively. There is a clear systematic 
trend with frequency, strongly supporting a frequency dependence of the nulling phenomenon, 
and suggesting that longer nulls are more common at lower observing frequencies.}
\label{fig:nullratio}
\end{figure}

In order to quantify this, we have computed individual null and burst durations 
from our binary time series of pulse states and the results are summarised in 
Fig.~\ref{fig:nullburst} in the form of histograms of null and burst lengths.  
The following conclusions can be drawn from these results: 
\begin{enumerate}
\item
This pulsar shows predominantly short duration nulls (up to four pulse periods). 
A vast majority of the observed nulls have durations of just one pulse period,
with a smaller fraction extending to durations of two or more pulse periods. 
\item
The number of longer duration nulls (i.e. two or more pulse periods) seems to
decrease with an increase in frequency.  This trend is very clear for data set I.
\item
The burst duration varies over a wide range from 1 to 30 pulse periods and 
seems to follow a roughly exponential-type distribution. Majority of these
bursts have durations of 1 to 10 pulse periods.
\item
Although there does not seem to be any systematic trend for burst durations
with frequency, short duration bursts are comparatively larger in number at 
our highest observing frequency of 4850 MHz.
\end{enumerate}
To further explore the frequency dependence of null durations, we study the
ratio of the number of nulls of single pulse duration, to the number of nulls of two, three
and four pulse durations.  Figure~\ref{fig:nullratio} shows these ratios for each
of the four frequencies.  It is clear from this that the relative number of longer
duration nulls does indeed decrease significantly with increasing frequency.  Thus
this pulsar appears to null more often for relatively longer durations at lower
frequencies.

\subsection{Selective Nulling: ``Exclusive Null'' Pulses}\label{s:exnulls}

In order to characterise the selective nulling phenomenon pointed out in section 
\S\ref{s:bin}, we compute the number 
of ``exclusive nulls" for every frequency combination -- these are the counts of
pulses that null {\it exclusively} for that frequency combination.  For instance, for the 
325+610+1400 MHz combination\footnote{To denote a specific frequency combination, we 
insert the ``+" symbol in between the relevant frequencies, e.g. the notation ``325+610 MHz" 
refers to the null pulses that are exclusive to 325 and 610 MHz.}, it is the difference 
between the null count for this combination and that for all four frequencies. For 
two-frequency combinations, we further exclude selective null pulses for the relevant 
3-frequency combinations (e.g., 325+610+1400 MHz and 325+610+4850 MHz for the case
of 325+610 MHz).  
Columns (3) and (5) in Table~\ref{tab:nf} tabulate such exclusive null counts (and 
their percentages). 
 
As per Table~\ref{tab:nf}, the most striking case among the various classes of selective 
nulling is a significantly large number of exclusive null pulses for the frequency 
combination 325+610+1400 MHz (3.5\% and 7.5\% respectively for \dataa and \datab). 
The combination 325+610 MHz also shows a 
significant number of exclusive nulls (2.7\% and 1.5\% for \dataa 
and \datab respectively).  In most other cases (barring the peculiar cases of 610+1400 
MHz for \datab, which we address in Appendix A), the number of such exclusive nulls 
is relatively small ($\la$ 0.5\%), and a careful examination of the relevant pulses do 
not show evidence for any general trend.  Therefore, in the bulk of the remainder of 
this paper, we focus our attention on the most interesting case of simultaneous 
exclusive nulls at 325+610+1400 MHz. In order to distinguish this special class of 
nulls from those that are broadband over the full range of observing frequencies, we 
refer to them as ``low frequency nulls'' (LF nulls). 

\section{Selective Nulling: Low Frequency Nulls at 325, 610 and 1400 MHz}\label{s:lfbn}

\begin{figure}[]
\centering
{\includegraphics[width=8cm]{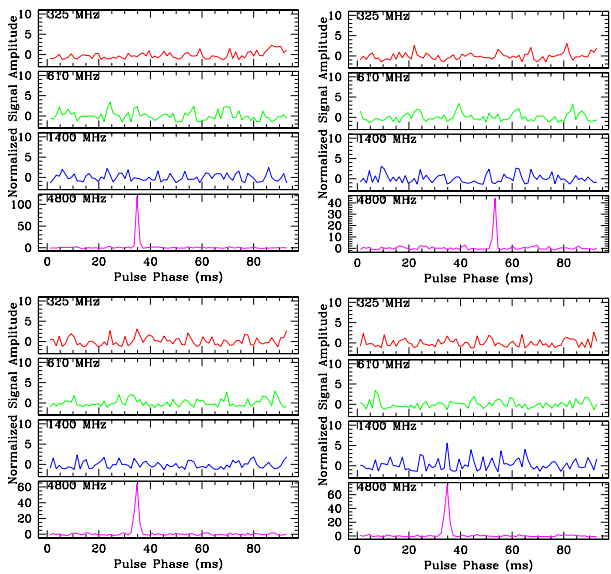}}
\caption{Examples of low frequency nulls which are marked by a strong, narrow
emission feature at the highest frequency of observation, 4850 MHz. These 
pulses are seen with a signal-to-noise ratio (S/N) as much as 300. The pulse 
phase window shown in these plots spans approximately 28 degree (92.8 ms) 
centred at the centre of the integrated pulse profile. The pulse intensities 
are normalised with off-pulse noise rms after subtracting the mean off-pulse 
level. No detectable emission seen at the three lower frequencies while the 
signal is detected with a S/N of 50 to 100 at 4850 MHz.}
\label{fig:eg}
\end{figure}

We have carefully examined the individual pulses that are grouped as
``low frequency nulls," and find, quite interestingly, that the emission at the highest 
frequency (4850 MHz) is often marked by a fairly strong, narrow pulse. 
Examples of this kind of pulses are shown in Fig.~\ref{fig:eg}.  In this
section, we characterise their properties, and compare and contrast them 
with those of normal emission seen at this frequency. 

\subsection{Characterisation of Emission at 4850 MHz}\label{s:char}

\begin{figure}[b]
\resizebox{\hsize}{!}{\includegraphics[]{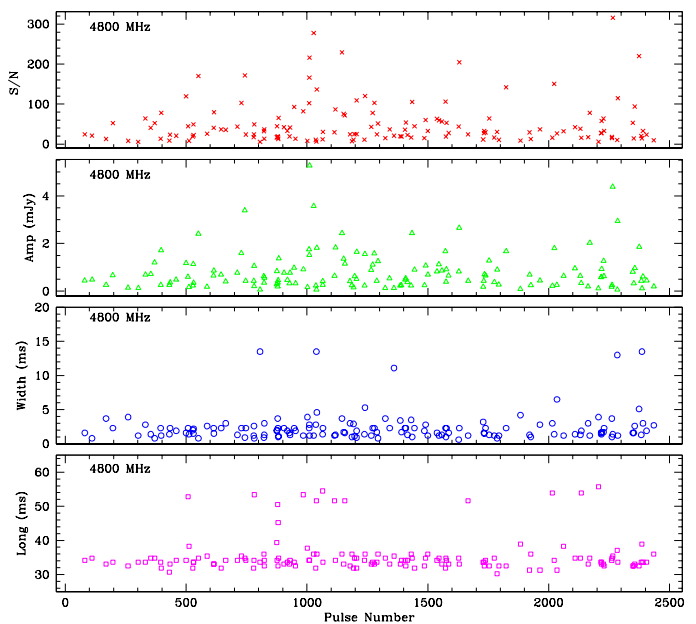}}
\caption{Properties of the narrow emission features seen at 4850 MHz 
during the times of low frequency nulls. The quantities plotted here 
are (from top to bottom) the peak signal-to-noise (S/N), amplitude, 
width and location (longitude) of these emission features. The results 
are for the \datab.}
\label{fig:gaus}
\end{figure}

To characterise the high frequency pulses occurring during the LF nulls, we use basic 
properties such as their width, location and strength.  This is done in a two-step 
process. First, we perform a box car analysis in order to obtain some first order 
estimates of these quantities. For this, we trial a large number of box car widths
(within a range from 0.1 to 0.9 times the main pulse width), 
deriving the amplitude, location and peak signal-to-noise ratio (S/N) in each case. 
The case that yields maximum S/N is then taken as the closest representation of the 
pulse.  Using these values as starting points ensures a quick and easy convergence of 
the finer grid search that is performed subsequently.  Here, the amplitude, width and 
location of the best fit Gaussian is determined using standard chi-square minimisation 
techniques. These three parameters, along with the peak S/N, are presented in 
Fig.~\ref{fig:gaus}.  From this, we can infer the following about the nature of these
high frequency emissions:

\begin{enumerate}
\item
The emission features are often quite strong, with peak S/N typically $\sim$50, 
while stronger pulses are seen with S/N as large as $\sim$300.
\item
Their occurrence is largely confined to a narrow longitude range of $\la$ 10 ms, 
starting roughly 0.25$w_p$ from the leading edge of the pulse profile, where 
$w_p$ is the ON pulse window.
\item
There is some evidence for a second preferred location (roughly at 0.75$w_p$ from the 
leading edge), albeit this is mainly seen in \datab. 
\item
These pulses typically have widths of a few milliseconds, and are usually narrower than 
the typical subpulses seen during the normal emission.
\end{enumerate}

Histograms of the above-mentioned parameters are plotted in Fig.~\ref{fig:full} 
(shaded regions). 
These are for \datab, however similar trends are also seen for \dataa. The S/N and 
amplitude seem to follow somewhat exponential-like distributions on a linear scale, 
but we have chosen to plot them on a logarithmic scale in order to better compare 
and contrast them with the corresponding distributions for the normal population 
of pulses (see next section).  The width distribution is somewhat skewed and is 
slightly double-peaked in nature.  The longitude distributions are nearly symmetric. 

\subsection{Comparison with ``Normal" Emission at 4850 MHz}\label{s:comp}

\begin{figure}[t]
\resizebox{\hsize}{!}{\includegraphics[]{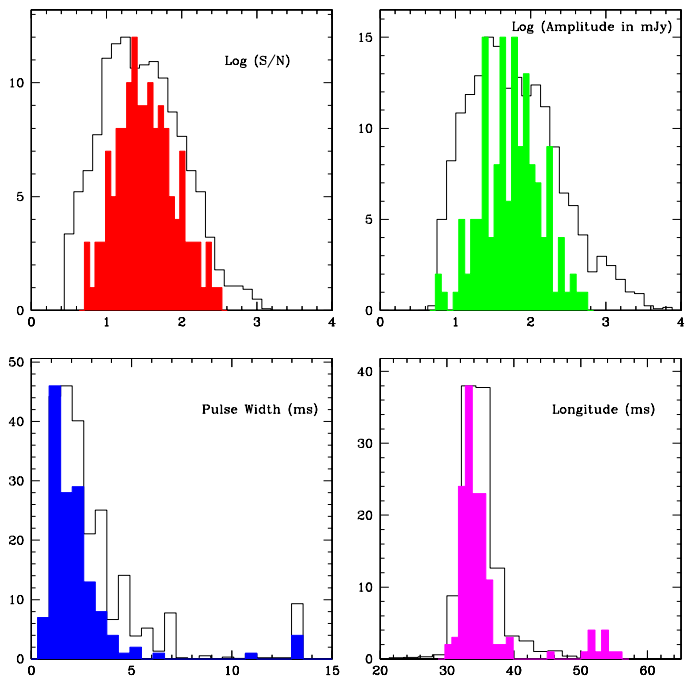}}
\caption{Properties of the high frequency (4850 MHz) emission during LF nulls 
(shaded regions) are compared with those of the {\it normal} emission (for the 
leading component only) at this frequency. These results are for the \datab. 
The top panels are the histograms of S/N and amplitude, with data binned at 
logarithmic intervals.  The binning is on a linear scale for the width and 
longitude distributions (bottom panels).}
\label{fig:full}
\end{figure}

In order to compare the properties of these high frequency (HF) pulses with 
those of ``normal'' pulses, we compute similar parameters for the sample of 
normal pulses at 4850 MHz. Normal pulses are those which do not belong to 
either the LF null class or the all-frequency null class. 
Since this pulsar shows a two-component profile, and given that the 
high frequency emission during LF nulls tends to occur preferentially in the 
longitude range of the leading component, we find that it is more appropriate 
to compare with the properties derived from the leading component only.  
The results of this exercise are summarised in the histograms shown in Fig.~\ref{fig:full} 
(for \datab), where quantities for the normal population of pulses are shown as 
unshaded bars.  For the reasons described earlier, 
the S/N and amplitude histograms are plotted on a logarithmic scale. Further, the two 
distributions are normalised for equal peaks to allow an easy comparison. 

\begin{figure}[b]
\centering
{\includegraphics[width=4cm]{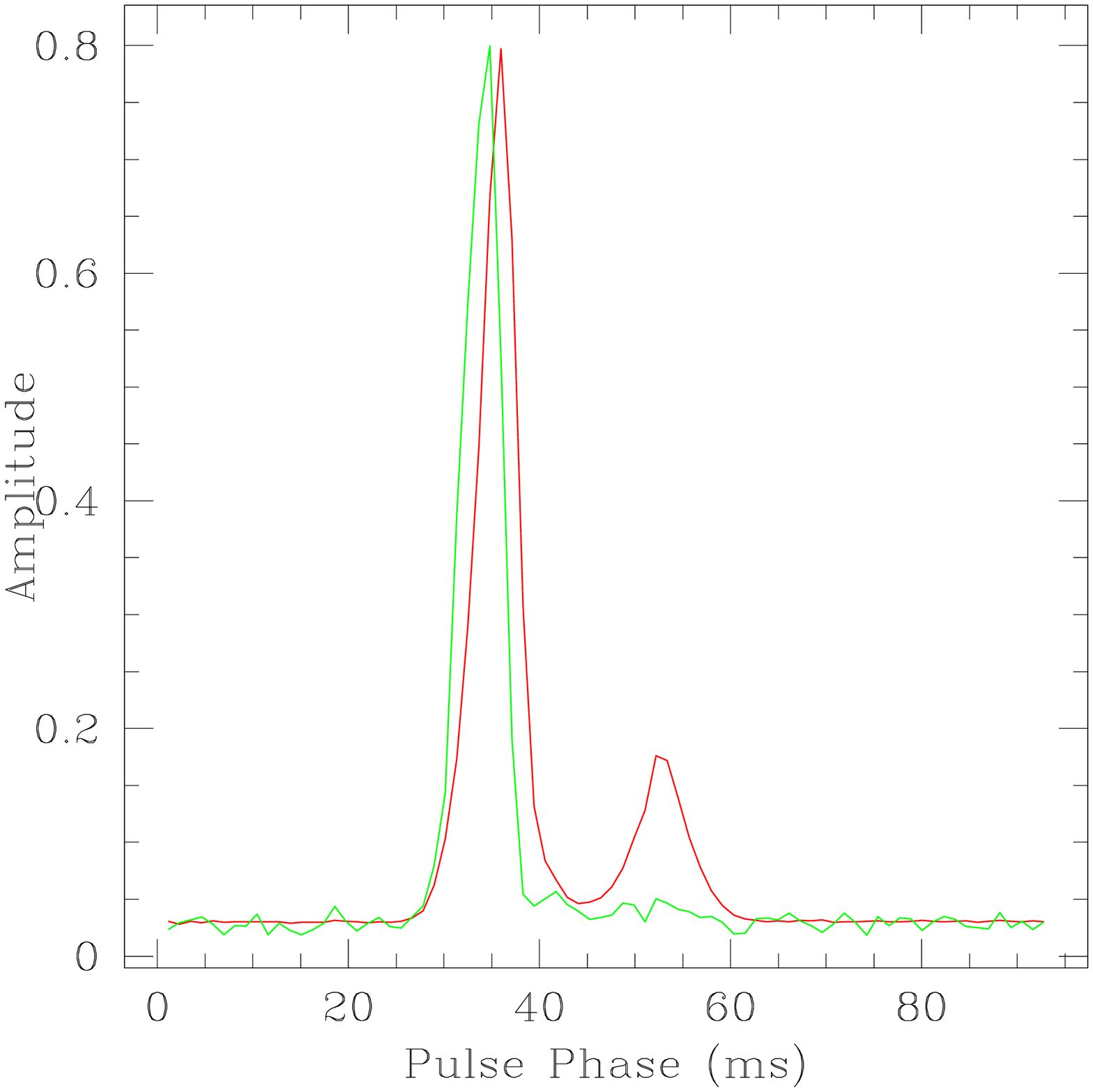}}
{\includegraphics[width=4cm]{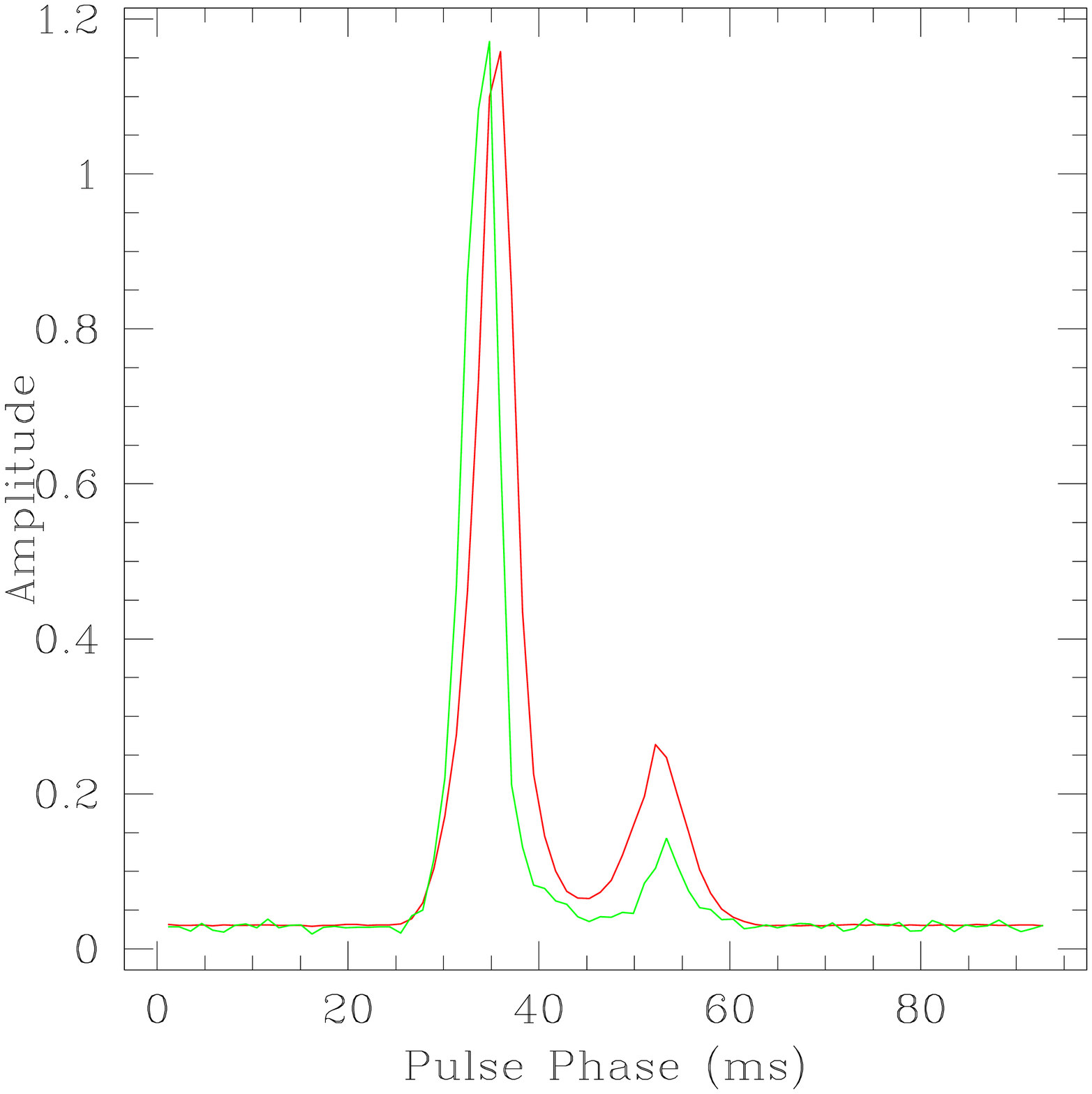}}
\caption{Average profiles of the emission at 4850 MHz during the low
frequency nulls (green curves) overlayed on those of the normal pulses
(red curves); {\it left}: the \dataa, and {\it right}: the \datab.
The peaks of the two profiles are slightly offset, indicating a 
tendency for the pulses during low frequency nulls to arrive at 
a slightly earlier phase.}
\label{fig:profs}
\end{figure}

From these figures we see that the S/N distribution is narrower and less 
skewed for the the HF emission during LF nulls and it appears to peak at 
a comparatively higher S/N. The amplitude distributions are also narrower and 
more symmetric than those for normal pulses. 
The distributions for widths and locations also show significant differences 
between the two classes.  As inferred qualitatively earlier, the width distribution 
for the HF pulses is narrower than that for the normal pulses.
There is significant spread in the longitudes of occurrence (for 
both classes), and the spread is more for the normal pulses, thus reinforcing 
the conclusion that the HF pulses are confined to a narrower range of longitude. 
Furthermore, the longitude distributions for the LF null class peak at slightly 
earlier longitudes than those for the normal pulses, indicating that this emission 
arrives at an earlier phase than that of the normal pulses.
Some of these results are clearly seen in a comparison of the average profiles 
obtained for the two classes of pulses (Fig.~\ref{fig:profs}).  These reinforce
the conclusions that the HF pulses during LF nulls occur predominantly under the
leading component and arrive at an earlier longitude than the normal pulses. 

\subsection{Comparison with ``Giant Pulses'' at 4850 MHz}\label{s:giant}

A different analysis of this data set in Paper IV revealed an interesting sub-class 
of ``giant'' pulses for this pulsar at 4850 MHz.  These are some of the strongest 
pulses seen among the non-nulling detections, with typical flux densities more 
than $ 10 \times \avS $ (where \avS is the mean flux density of the entire data), 
which is the commonly used threshold for detection of giant pulses 
(e.g. \citet{johnston2002}).  The emission is often narrow in width and tends to 
occur preferentially at the {\it trailing} edge of the leading pulse component 
(see Fig. 17 of Paper IV).  Based on their large flux densities and from the hint 
of an emerging power-law component in the cumulative distribution of their energies 
(Fig. 16 of Paper IV), we referred to them as possible giant pulses. 
However, unlike the classical giant pulses as have been observed for the Crab 
and a few other pulsars \citep{cordes2004,johnston2002}, which are characterised 
by a power-law energy distribution and a spectral index that is comparable to or
steeper than that of the normal pulses, these appear to be of a significantly 
flatter spectrum. This conjecture stems largely from (i) their apparent absence 
at our lower frequencies and (ii) from observations at a higher frequency of 8450 
MHz (Maron \& L\"ohmer, private communication).

It is important to note that in Paper IV we studied the same data set but excluded 
all pulses with nulls at any frequency in order to compute the spectra. Hence, when
studying the nulling pulses and their frequency behaviour as we do in this paper,
it is the study of a different subset of pulses. Given this, it is interesting 
to compare and contrast the population of the afore-mentioned giant pulses with the 
HF pulses, as seen during the LF nulls.  Both classes are generally strong and narrow, 
with a mostly single-peaked emission, though the flux densities of the latter are 
well below the $ 10 \times \avS $ threshold of the former. Further, they both tend 
to occur preferentially over some restricted longitude ranges of the leading component: 
giants near the trailing edge of the leading component (i.e. arriving at a slightly 
later phase than the normal pulses), and the HF pulses near the leading edge (arriving 
at a slightly earlier phase than normal pulses).  Both show low-number statistics (40 
giants and 195 HF pulses) which prevent us from a detailed study of their energy 
distributions.  Their rates of occurrence -- typically 1 in 100 for giants and 
1 in 20 for HF pulses -- are comparable or even larger than the rates of occurrence 
of giant pulses that are seen for the Crab pulsar \citep[e.g.][]{cordes2004}.  Thus 
there are some similarities between these two special classes of pulses, and it is 
possible that they may have related origins.

\subsection{Selective Nulling at 325 and 610 MHz}
\label{s:exnullgmrt}

\begin{figure}[b]
\centering
{\includegraphics[width=8cm]{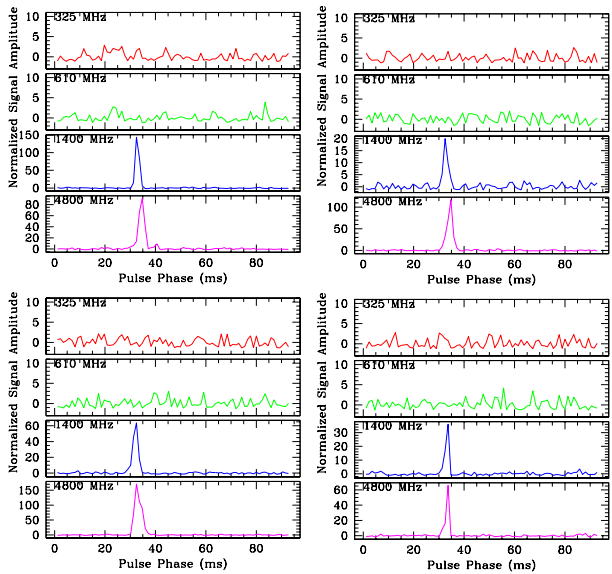}}
\caption{Examples of pulses that null at the lower two observing frequencies, 
325 and 610 MHz, while the emission is seen -- mostly as narrow, single-peaked 
pulses -- at the two higher frequencies, 1400 and 4850 MHz. Statistics of this 
class of pulses is however not good enough to construct meaningful histograms 
such as in Fig.~\ref{fig:full} (see text). The pulse intensities are normalised 
in a manner similar to that in Fig.~\ref{fig:eg}. No detectable emission is 
seen at 325 and 610 MHz even upon averaging all the null pulses at these 
frequencies.}
\label{fig:exnulllow}
\end{figure}

The next most interesting sample in our data set is a subset of pulses
that null {\it exclusively} at 325 and 610 MHz. Some examples are shown 
in Fig.~\ref{fig:exnulllow}.  Although an analysis similar to that described in 
\S~\ref{s:char} and \S~\ref{s:comp} was attempted for this class of nulls, the poor 
statistics of the sample did not allow us to obtain any meaningful characterisation 
of the emission at higher frequencies for this class of LF nulls.  Qualitative 
examination shows that a majority ($\sim$80--90\%) of these pulses are characterised 
by single-peaked emission at 1400 and 4850 MHz, although in most cases the pulse 
widths do not seem to be significantly narrower than typical sub-pulse widths. As in
the case of 
the first class of HF pulses, the emission here (at 4850 MHz) also tends to occur 
preferentially at the longitude of the leading component, with only a minority of 
these pulses appearing at the longitude of the trailing component.
Only occasionally, the emission at 1400 and/or 4850 MHz is seen as a double-peaked 
pulse that is characteristic of the normal emission from this pulsar. 

\begin{figure}
\centering
{\includegraphics[width=4.25cm]{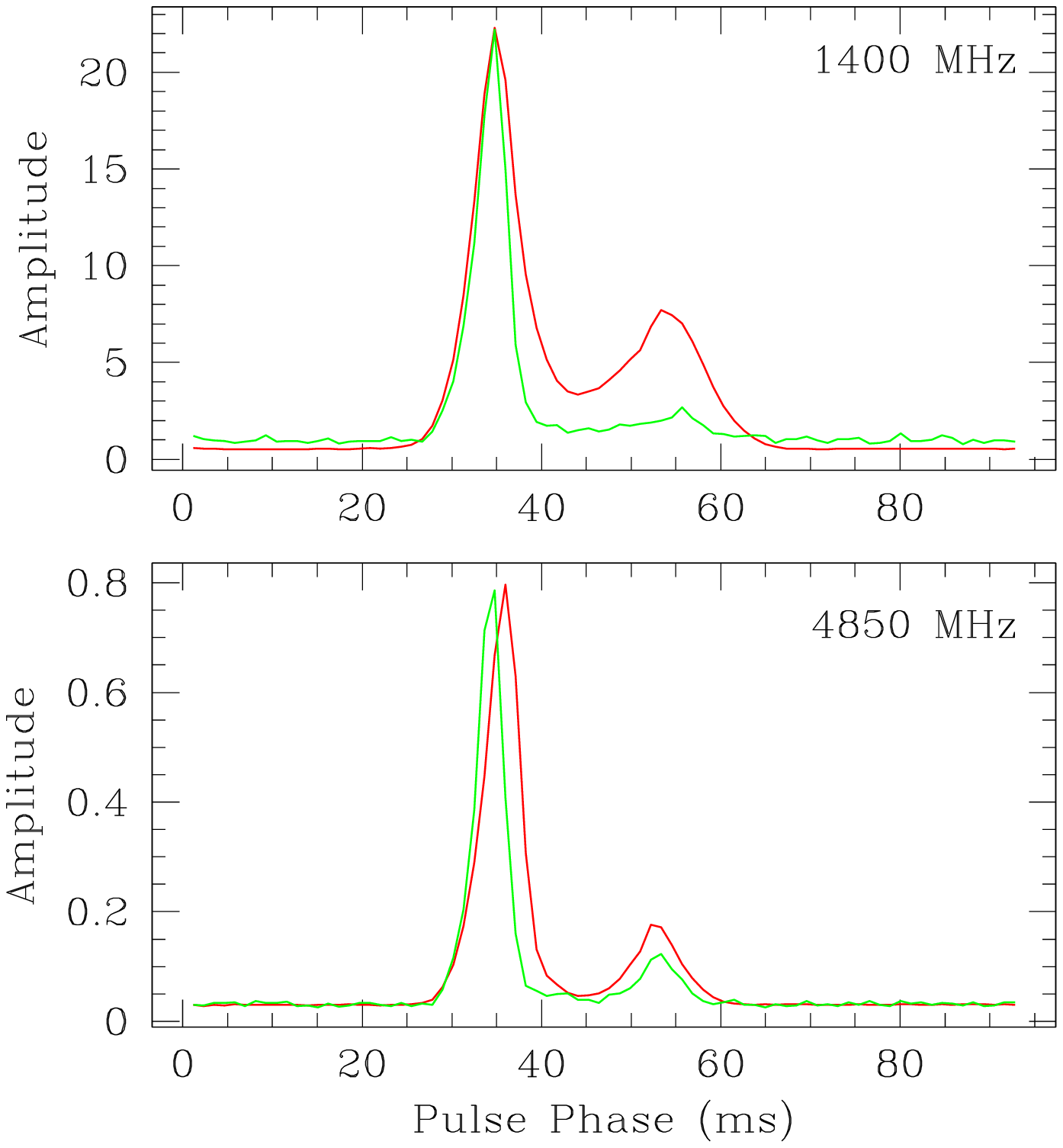}}
{\includegraphics[width=4.25cm]{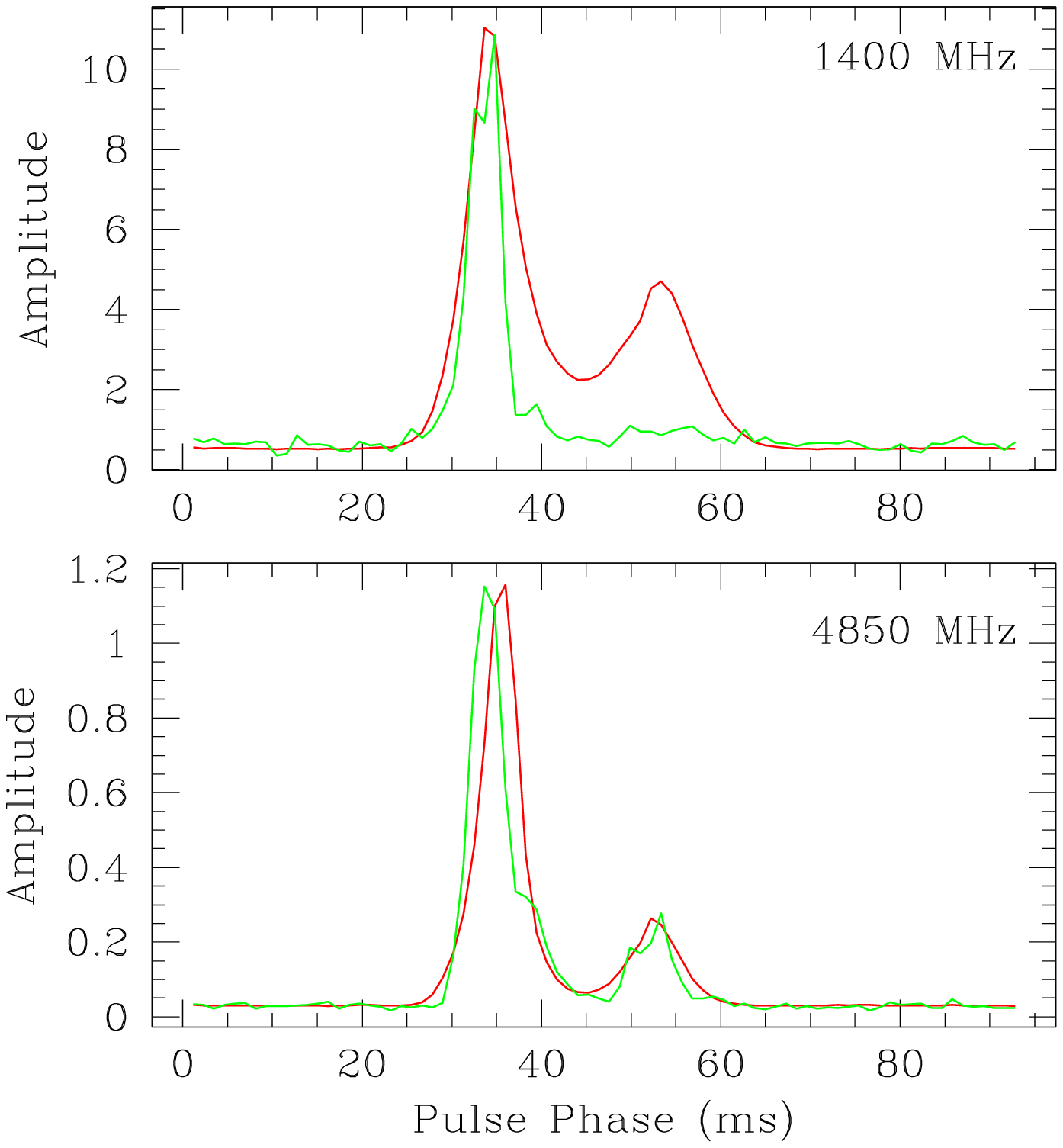}}
\caption{Average profiles constructed from the subsets of pulses where
the emission is seen at 1400 and 4850 MHz, while the pulsar switches to 
a ``null state'' at 325 and 610 MHz (green curves). Overlayed are the 
average profiles constructed from the sample of normal pulses at these 
frequencies (red curves). The emission at both frequencies tends to
occur preferentially at the longitude of the leading pulse component; 
the profiles are nearly peak-aligned at 1400 MHz (top panels), while a 
slight phase offset is evident for those at 4850 MHz (bottom panels). 
The profiles of the subsets are scaled for a close match with the peaks 
of the average profiles of the normal pulses. The left and right panels 
correspond to the data sets I and II respectively.}
\label{fig:2freqprofs}
\end{figure}

The average profiles (at 1400 and 4850 MHz) of the pulses in this null class are shown 
in Fig.~\ref{fig:2freqprofs} (green curves), along with those of the full sample of 
normal pulses (red curves). As was the case for the first class of HF pulses,
these pulses tend to arrive at a slightly earlier phase than the normal emission, at 4850 MHz.
However, no such offset is readily visible for the emission at 1400 MHz.  





\section{Discussion}\label{s:res}

In this paper, we focus on the analysis of the pulse nulling phenomenon in PSR B1133+16 
using high-quality single-pulse data from simultaneous multifrequency observations.
The most important finding from our analysis is that nulling {\it does not always occur 
simultaneously} at all frequencies. Our observations uncover a significant number of 
pulses ($\approx$6\%) that null at the three lower frequencies of our observation, however
are marked by quite narrow and strong pulses at the highest frequency of 4850 MHz. Their 
properties seem to be quite different from those of the normal pulses seen at this 
frequency, but interestingly, show some striking similarities with the subset of the 
strongest pulses in such a sample. Our analysis shows some evidence for the number 
of simultaneous null pulses to decrease with the frequency coverage of the data, and 
also suggests that longer-duration nulls are relatively more common at the lower observing 
frequencies.  In the remainder of the paper, we review what has been learnt so far 
about the nulling behaviour of this pulsar from past observations and discuss possible 
implications for pulsar emission.

\subsection{Comparison with Previous Nulling Studies}\label{s:prev}

PSR B1133+16 is among the well-studied pulsars since the early days 
of pulsar observations 
\citep{backer1972,ferguson1978,kardashev1982,boriakoff1983,smirnova1994}. 
Its nulling behaviour is not extreme as in the case of PSR B0031$-$07 
and PSR B1944+17 \citep{hugu1970,deich1986}, which are known for their 
long durations of nulling (null fractions of the order $\sim$50\%).
In fact, most null durations of this pulsar are within the range of one 
to a few pulse periods, with a moderately large null fraction 
($\sim$10 to 20\%) over typical observing durations.
It is also important to recognise that most nulling studies to date
have been based on data taken at a single observing frequency.
Interestingly, our estimates of null fraction at single frequencies (with
the exception of 4850 MHz for \datab and barring the special cases of 610
and 1400 MHz for \datab; see Appendix A), are comparable to the published
estimate of 15$\pm$3\% from earlier studies (Biggs 1992; Ritchings 1976).
However, the null fraction of real ``broadband'' nulls (i.e., nulls that
simultaneously occur at all four frequencies of observation) is only
$\approx$7.5\%. Thus, in general, most published estimates for null
fraction are likely to be overestimates if broadbandness is adopted as
an additional criterion to define a truly null state.

There have been several single-pulse studies of
PSR B1133+16 in the past based on data from
simultaneous observations at multiple frequencies
\citep{boriakoff1983, kardashev1982,
boriakoff-ferguson1981,bartel-sieber1978,backer1974},
and it is fairly well established that the pulse
energy and structure of this pulsar are well
correlated at single-pulse, sub-pulse and micro-pulse
levels. Specifically, the works of \citet{boriakoff1983}
and \citet{kardashev1982} addressed the broadbandness
of micropulse and subpulse structure, while
\citet{bartel-sieber1978} focussed on the microstructure
of single pulses in this pulsar. However, none of these studies
were able to address the simultaneity in pulse nulling at different
observing frequencies. In fact, there have been few studies
that have even peripherally addressed this interesting phenomenon.
The only relevant work that we are
aware of in this context is by \citet{davies1984}
who studied simultaneous data of PSR B0809+74 taken at
102 and 406 MHz. In their short stretch of data (348
pulse periods), they recorded 9 nulls at 102 MHz,
and only 3 at 406 MHz. Interestingly nulls at 406 MHz
always correspond to nulls at 102 MHz,
but not vice versa. While this may not qualify as a
strong supporting evidence for selective nulling, it
does suggest that the pulsar nulling characteristics
vary with observing frequency.

\subsection{Implications for Pulsar Emission Models}\label{s:impli}

In sections \S\ref{s:char} and \S\ref{s:comp}, we compared and
contrasted the characteristics of the high frequency emission
during low frequency nulls with the ``normal'' emission at 4850 MHz.
We also compared them with those of ``giant'' pulses as reported 
in our earlier paper (\S\ref{s:giant}).
Based on our analysis, we would like to conjecture that  
(a) the high frequency emission pulses are different from the normal
pulses and are likely to be related to the ``giant'' pulses; 
(b) the spectrum of these ``giants'' is probably flatter, and 
consequently they are more easily seen at higher frequencies 
(Paper IV); and (c) we see them when the normal radio emission is 
off or too weak to be detectable (i.e. nulling) at high frequencies 
(this paper).

One of the most interesting results from our analysis is the observation
of narrow high frequency pulses at times of low frequency nulls.
The simplest possible interpretation could be the presence of
an {\it additional} process of emission that does {\it not} 
turn off when the pulsar switches to a ``null state'' at low 
frequencies.  It is also quite possible that such a process is 
more like a gradual effect, becoming predominant above a cutoff 
frequency. Unfortunately, our data are insufficient to address 
such a hypothesis. 

\begin{figure}
\centering
{\includegraphics[width=8cm]{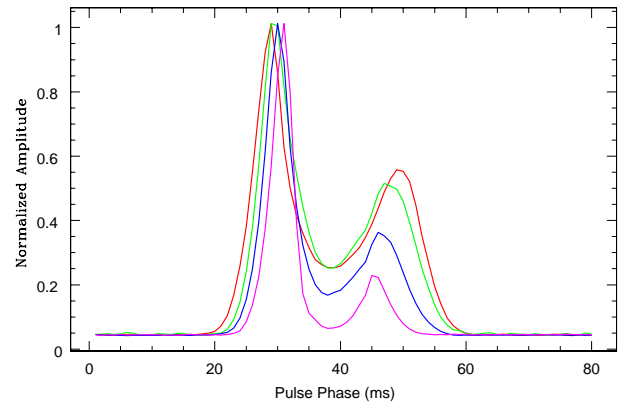}}
\caption{Integrated pulse profiles at all four frequencies of observation; 
325 MHz (red), 610 MHz (green), 1400 MHz (blue), 
and 4850 MHz (magenta). 
These profiles are constructed from data that are already time-aligned 
and corrected for frequency-dependent dispersive delays.
Further, they are normalised for equal peaks of the leading pulse 
components. The relative strength of the leading pulse component 
shows a systematic increase with an increase in frequency.}
\label{fig:allprofs}
\end{figure}

Our data also show several instances of significant emission at 
both 1400 and 4850 MHz, when there occurs a null at the lower 
frequencies. However, the results are not conclusive enough to 
suggest a frequency dependence for such an additional process.  
A likely scenario is that the emission tends to be narrower and 
peaks at an earlier phase with an increase in frequency.  

It is possible that such an additional emission process is also present
during the normal emission. However, given that the pulsar's emission is 
typically double-peaked and has a strong leading pulse component, its
presence might be hard to discern in non-nulling pulses.
Assuming this may be the case, we would expect a significant increase 
in relative strength of leading and trailing pulse components with an 
increase in frequency, which is indeed seen for this pulsar 
(Fig.~\ref{fig:allprofs}).

An alternate, and perhaps more likely, possibility involves this
emission being related to the population of ``giant'' pulses (as
discussed in \S~\ref{s:giant}). In particular, given their
preferred longitudes of occurrence near the leading pulse
component, a likely flat-spectrum nature, and above all, a tendency
to occur even when the pulsar nulls at low frequencies, may also
suggest that they probably originate in a different part of the
magnetosphere which does not participate in nulling. If this is
the case, it may support an outer gap origin as opposed to a
polar cap region (where nulling is probably more effective).
Such a scenario might also suggest a possible link to some
high-energy emission processes that occur in the outer parts
of the magnetosphere. Future simultaneous observations of
this pulsar, preferably with a denser sampling of frequency in
the $\sim$1--5 GHz range, should enable us to address this
in more detail.

\section{Conclusions}\label{s:conc}

We have studied high-quality single-pulse data of PSR B1133+16 from
simultaneous multifrequency observations conducted using the
GMRT, Lovell and Effelsberg to perform an in-depth analysis of
the pulse nulling phenomenon. Our observations provided long data
stretches over a wide frequency range (from 0.3 to 4.85 GHz),
which allow, for the first time, investigation of nulling as
a function of observing frequency, separation in frequency as
well as combination of frequencies. The data shows that the pulsar
spends approximately 15\% of the time in a null state at our
frequencies of observation.
However, only roughly half of these nulls occur simultaneously at
all four frequencies.  In other words, much in contrary to the
traditional notion of being a broadband phenomenon, nulling does
{\it not always} occur simultaneously at all four frequencies.

Our most interesting finding of this ``selective
nulling'' phenomenon is a significantly large number of pulses
($\approx$5\%) that show an emission at the highest frequency
of our observation, 4850 MHz, while there occurs a null at all
three lower frequencies (325, 610 and 1400 MHz).
This emission is often seen as fairly strong and narrow
pulses, and is different from the broad, double-peaked emission
normally seen from this pulsar.  We have characterised these high
frequency pulses in terms of their amplitudes, widths and locations,
and compared their statistical properties with those of the leading
component of the normal pulses.
Our analysis reveals significant differences in the properties
between the two classes. Specifically, the population of HF
pulses shows an amplitude distribution that is more skewed,
a comparatively narrower width distribution, and a tendency to
arrive at a slightly earlier phase than the leading pulse
component. Additionally, we note some interesting similarities
between these narrow pulses and the ``giant'' pulses as
identified for this pulsar in our earlier paper.

We do not have any clear and convincing explanation to offer
at this point, apart from the simplest (and rather speculative)
interpretation involving the presence of an additional emission
process that does not turn off even when the pulsar nulls at
lower frequencies. Alternatively, such an additional process may
be a gradual effect that becomes more predominant at higher
frequencies. While our data are insufficient to address such
hypotheses, such scenarios could potentially result in
observable effects such as an increase in relative strength
between the leading and trailing pulse components at higher
observing frequencies, which is indeed seen for this pulsar.

Given the significant similarities between this emission and
``giant'' pulses as seen at 4850 MHz, it is quite possible
that they share a common, or at least related, origin.
In particular, their tendency to occur at specific pulse
longitudes, along with the pulse shapes and spectral
characteristics that are different from the normal emission,
add support to such a conjecture and are suggestive of a likely
origin in the outer parts of the magnetosphere. We hope
future simultaneous observations will help to shed more
light on such possibilities.

\medskip

\noindent
{\it Acknowledgements:}
We are grateful to all the people working at the participating
telescopes who made these observations possible. In particular
we thank S. Kudale, M. Jangam and C. Lange for assistance with
the observations, and R. Wielebinski for the encouragement and
support towards this project. We also thank an anonymous referee 
for a critical review and several insightful comments which 
helped improve the contents and presentation of the paper.
The Giant Metrewave Radio Telescope (GMRT) is operated by the 
National Centre for Radio Astrophysics of Tata Institute of 
Fundamental Research.


\bigskip

\bigskip


\noindent
{\bf Appendix A: Scintillation Fading and Nulling Statistics}\label{s:scnt}

\bigskip

Here we address the peculiar case of a large number of nulls seen at 610 
and 1400 MHz for \datab.  On a careful examination, we find that a majority 
of these nulls are largely clustered into two groups, the first in the range 
$\sim$1200 to $\sim$1500 in pulse number, and the second in the range 
$\sim$2200 to $\sim$2500.  Interestingly, these regions of occurrence correlate 
well with the regions of flux fading seen at these frequencies. We therefore 
propose that the apparent excess in null counts and simultaneous nulls 
exclusive to these frequencies is essentially a manifestation of  
scintillation-induced flux fading. Below we elaborate on our arguments.

As is well known, pulsar signals are subject to interstellar scintillation (ISS),
observable effects of which are strongly frequency dependent.
At low observing frequencies, most pulsars are in the strong scintillation
regime, where the observable effects can be grouped into two distinct 
classes, viz., diffractive and refractive scintillation 
\citep[e.g.][]{rickett1990,cordes1986,romani1986}, with consequences
such as flux density variations exhibiting two characteristic time scales. 
At frequencies higher than a few GHz, most nearby pulsars transition 
to the weak scintillation regime, which is marked by a single characteristic 
time scale \citep{rickett1990}. PSR B1133+16 is expected to be in the strong 
scattering regime at 325, 610 and 1400 MHz, and in weak scintillation at our
highest observing frequency of 4850 MHz (see Paper IV for details on relevant
scintillation parameters).

For observations at low observing frequencies ($\la$ 1 GHz), 
the dominant ISS effect is random intensity modulations due to
diffractive scintillation \citep[e.g.][]{bhat1999}. The pulse intensity decorrelates 
over narrow ranges in time and frequency, and the decorrelation widths 
($\nud$ and $\tiss$ respectively in frequency and time) are strongly 
frequency dependent ($\nud \sim \nu^{4.4}$, $\tiss \sim \nu^{1.2}$, 
where $\nu$ is the frequency of observation). Thus, intensity modulations 
as deep as 100\% can be expected over time and frequency scales of $\tiss$ 
and $\nud$. For observations made over time durations (\tobs) much 
larger than $\tiss$, three regimes may be identified for apparent 
intensity modulations, depending on the ratio of the scintillation 
bandwidth ($\nud$) to the total bandwidth of observation, $\delB$:
(i) $\delB \gg \nud$: here the effective flux modulations will be quenched to 
some extent depending on the ratio $\delB/\nud$; (ii) $\delB \sim \nud$, i.e.
when the observing bandwidth is comparable to the scintillation bandwidth: 
flux fading as large as $\sim$100\% is likely to occur over durations of $\tiss$;
(iii) $\delB \ll \nud$, such deep flux fading could still occur over $\tiss$, albeit 
with somewhat less likelihood. 

Consider the scenario where simultaneous observations made at three 
different frequencies correspond to the three different cases described 
above. There could be rare instances of simultaneous flux fading at the 
two frequencies that correspond to cases (ii) and (iii). Such apparent 
nulls may lead to consequences such as an increase in null counts as
well as an increase in the number of simultaneous nulls exclusive to 
those frequencies. A comparison of the scintillation and observing 
parameters for PSR B1133+16 at 610 and 1400 MHz ($\nud$ $\sim$11 MHz, 
$\tiss$ $\sim$ 6 min at 610 MHz; $\nud$ $\sim$ 270 MHz, 
$\tiss$ $\sim$ 16 min at 1400 MHz) supports such a conjecture.
 
Some simple checks could help reaffirm this. Since \tiss increases with the 
observing frequency, we would expect the time span of such apparent nulls 
to be limited to \tiss at the lower frequency, i.e. 610 MHz. Interestingly, the 
observed duration of ~300 pulses ($\approx$6 min) is in excellent agreement 
with the expected value for \tiss at 610 MHz. Further, \tiss at 1400 MHz is 
expected to be roughly 3 times larger than that at 610 MHz. This means we 
can expect null counts at these two frequencies to increase by roughly ~4\% 
and ~12\% in comparison to the measured values for the \dataa. Though 
the observed excess is somewhat larger than this, it does conform 
to the expectation that larger excess should be seen at 1400 MHz.
In short, the increase in null counts and a larger number of exclusive nulls
at 610 and 1400 MHz for \datab can be attributed to the combination of 
ISS and our observing parameters.


\begin{thebibliography}{}
\bibitem[Backer(1970)]{ba} 
\bibitem[Backer(1970)]{backer1970} Backer, D. C.\ 1970, Nature, 228, 42
\bibitem[Backer(1972)]{backer1972} Backer, D.~C.\ 1972, \apjl, 
174, L157 
\bibitem[Backer \& Fisher(1974)]{backer1974} Backer, D.~C.~\& 
Fisher, J.~R.\ 1974, \apj, 189, 137 
\bibitem[Bartel \& Sieber(1978)]{bartel-sieber1978} Bartel, N.~\& Sieber, 
W.\ 1978, \aap, 70, 307 
\bibitem[Bhat, Rao, \& Gupta(1999)]{bhat1999} Bhat, N.~D.~R., 
Rao, A.~P., \& Gupta, Y.\ 1999, \apjs, 121, 483
\bibitem[Bhat et al.(2001)]{bhat2001} Bhat, N.~D.~R., Karastergiou, A., 
Gupta, Y., Kramer, M., \& Lyne, A.~G.\ 2001, BAAS, 34, 569
\bibitem[Biggs(1992)]{biggs1992} Biggs, J.~D.\ 1992, \apj, 394, 574 
\bibitem[Boriakoff(1983)]{boriakoff1983} Boriakoff, V.\ 1983, \apj, 
272, 687 
\bibitem[Boriakoff \& Ferguson(1981)]{boriakoff-ferguson1981} Boriakoff, V.~\& 
Ferguson, D.~C.\ 1981, IAU Symp.~ 95: Pulsars: 13 Years of Research on 
Neutron Stars, 95, 191 
\bibitem[Cordes et al.(2004)]{cordes2004} Cordes, J.~M., Bhat, N.~D.~R., 
Hankins, T.~H., McLaughlin, M.~A., \& Kern, J.\ 2004, \apj, 612, 375 
\bibitem[Cordes, Pidwerbetsky, \& Lovelace(1986)]{cordes1986} 
Cordes, J.~M., Pidwerbetsky, A., \& Lovelace, R.~V.~E.\ 1986, \apj, 310, 
737 
\bibitem[Davies et al.(1984)]{davies1984} Davies, J.~G., Lyne, 
A.~G., Smith, F.~G., Izvekova, V.~A., Kuzmin, A.~D., \& Shitov, I.~P.\ 
1984, \mnras, 211, 57
\bibitem[Deich et al.(1986)]{deich1986} Deich, W.~T.~S., Cordes, 
J.~M., Hankins, T.~H., \& Rankin, J.~M.\ 1986, \apj, 300, 540 
\bibitem[Durdin et al.(1979)]{durdin1979} Durdin, J.~M., Large, 
M.~I., Little, A.~G., Manchester, R.~N., Lyne, A.~G., \& Taylor, J.~H.\ 
1979, \mnras, 186, 39 
\bibitem[Ferguson \& Seiradakis(1978)]{ferguson1978} Ferguson, 
D.~C., \& Seiradakis, J.~H.\ 1978, \aap, 64, 27 
\bibitem[Gupta, Gothoskar, \& Bhat(2002)]{gupta2002} Gupta, Y., 
Gothoskar, P., \& Bhat, N.~D.~R.\ 2002, IAU Symposium, 199, 369 
\bibitem[Huguenin et al.(1970)]{hugu1970} Huguenin, G.~R., 
Taylor, J.~H., \& Troland, T.~H.\ 1970, \apj, 162, 727 
\bibitem[Johnston \& Romani(2002)]{johnston2002} Johnston, S.~\& 
Romani, R.~W.\ 2002, \mnras, 332, 109 
\bibitem[Karastergiou et al.(2001)]{karas2001} 
Karastergiou, A., von Hoensbroech, A., Kramer, M., Lorimer, D.~R., Lyne, A.~G., 
Doroshenko, O., Jessner, A., Jordan, C. \& Wielebinski, R. \ 2001, \aap, 379, 270 
\bibitem[Karastergiou et al.(2002)]{karas2002} Karastergiou, A., Kramer, M., 
Johnston, S., Lyne, A.~G., Bhat, N.~D.~R., \& Gupta, Y.\ 2002, \aap, 391, 247
\bibitem[Karastergiou et al.(2003)]{karas2003} 
Karastergiou, A., Johnston, S., \& Kramer, M.\ 2003, \aap, 404, 325 
\bibitem[Kardashev et al.(1982)]{kardashev1982} Kardashev, N.~S., Nikolaev, N.~I., 
Novikov, A.~I., Popov, M.~V., Soglasnov, V.~A., Kuzmin, A.~D., Smirnova, T.~V., 
Bartel, N., Sieber, W. \& Wielebinski, R.\ 1982, \aap, 109, 340 
\bibitem[Kramer et al.(2003)]{kramer2003} Kramer, M., 
Karastergiou, A., Gupta, Y., Johnston, S., Bhat, N.~D.~R., \& Lyne, A.~G.\ 
2003, \aap, 407, 655
\bibitem[Lorimer et al.(1998)]{lorimer1998} Lorimer, D.~R., Jessner, A., 
Seiradakis, J.~H., Lyne, A.~G., D'Amico, N., Athanasopoulos, A., Xilouris, K.~M., 
Kramer, M. \& Wielebinski, R.\ 1998, \aaps, 128, 541 
\bibitem[Rankin(1986)]{rankin1986} Rankin, J.~M.\ 1986, \apj, 301, 901 
\bibitem[Rickett(1990)]{rickett1990} Rickett, B.~J.\ 1990, \araa, 
28, 561 
\bibitem[Ritchings(1976)]{ritchings1976} Ritchings, R.~T.\ 1976, \mnras, 176, 249 
\bibitem[Robinson et al.(1968)]{robinson1968} Robinson, B.~J., Cooper, B.~F.~C., Gardner, F.~F., et al.\ 1968, Nature, 218, 1143
\bibitem[Romani, Narayan, \& Blandford(1986)]{romani1986} Romani, 
R.~W., Narayan, R., \& Blandford, R.\ 1986, \mnras, 220, 19 
\bibitem[Smirnova et al.(1994)]{smirnova1994} Smirnova, T.~V., 
Tul'bashev, S.~A., \& Boriakoff, V.\ 1994, \aap, 286, 807 
\end{thebibliography}
\end{document}